\documentclass[%
prd,twocolumn,nofootinbib,superscriptaddress
]{revtex4-2}

\usepackage{preamble}
\usepackage{bm}
\usepackage{comment}
\usepackage{subfigure}
\usepackage{lineno}

\begin{document}

\title{A Method for Testing Diffusive Shock Acceleration and Diffusion Propagation of 1-100 TeV Cosmic Electron with Multi-wavelength Observation of Geminga Halo and Pulsar Wind Nebula}

\author{Weikang Gao}
\email{gaowk@ihep.ac.cn}
\affiliation{Key Laboratory of Particle Astrophysics, Institute of High Energy Physics, Chinese Academy of Sciences, 100049 Beijing, China}
\affiliation{University of Chinese Academy of Sciences, 100049 Beijing, China}
\affiliation{Tianfu Cosmic Ray Research Center, 610000 Chengdu, Sichuan,  China}

\author{Li-Zhuo Bao}%
\email{baolizhuo@bjp.org.cn}
\affiliation{Beijing Planetarium, Beijing Academy of Science and Technology, 100044 Beijing, China}

\author{Kun Fang}
\affiliation{Key Laboratory of Particle Astrophysics, Institute of High Energy Physics, Chinese Academy of Sciences, 100049 Beijing, China}
\affiliation{Tianfu Cosmic Ray Research Center, 610000 Chengdu, Sichuan,  China}
\author{En-sheng Chen}
\affiliation{Key Laboratory of Particle Astrophysics, Institute of High Energy Physics, Chinese Academy of Sciences, 100049 Beijing, China}

\affiliation{Tianfu Cosmic Ray Research Center, 610000 Chengdu, Sichuan,  China}
\author{Siming Liu}
\affiliation{The School of Physical Science and Technology, Southwest Jiaotong University,  611756 Chengdu, Sichuan, China}
\author{HongBo Hu}
\affiliation{Key Laboratory of Particle Astrophysics, Institute of High Energy Physics, Chinese Academy of Sciences, 100049 Beijing, China}
\affiliation{University of Chinese Academy of Sciences, 100049 Beijing, China}
\affiliation{Tianfu Cosmic Ray Research Center, 610000 Chengdu, Sichuan,  China}


\date{\today}

\begin{abstract}

Diffusive shock acceleration and diffusion propagation are essential components of the standard cosmic ray model. These theories are based on extensive observations of high-energy solar processes, providing substantial direct evidence in the MeV energy range. Although the model is widely and successfully used to explain high-energy cosmic phenomena, direct validation has been elusive. The multi-wavelength spectra and angular profile measurements of the Geminga pulsar wind nebula and its pulsar halo, particularly the precise spectral observations by HAWC and LHAASO-KM2A in recent years, offer a rare opportunity to test these theories with cosmic rays energies between 1 TeV and several hundred TeV. These observations are expected to elevate the direct testing of theoretical models from multi-MeV to sub-PeV energies. In this work, a method is developed to test the diffusive shock acceleration and diffusion propagation model between one and several hundred TeV energies through the latest spectral and morphological data of the Geminga region from HAWC and Fermi-LAT.  
Our results show that the theories of diffusive shock acceleration and diffusion propagation are consistent with experimental observations. 
However, the published morphological data adopted rather wide energy bins and currently do not allow a high precision test of the inferred energy dependent diffusion coefficient by observed energy spectra with DSA theory. It is anticipated that future HAWC and LHAASO-KM2A observations will yield higher-precision results, and the confirmation of a rapidly increasing diffusion coefficient above 100 TeV would serve as important evidence supporting the diffusive shock acceleration and diffusion propagation theory.
We emphasize that such a confirmation specifically tests this acceleration and propagation theory, without confirming or excluding other models. Similar tests would be both important and valuable for other models.

\end{abstract}

\maketitle


\section{\label{sec1}Introduction}

As described in the review \cite{amato2024particle}, the acceleration mechanisms in pulsar wind nebulae (PWNe) include diffusive shock acceleration (DSA) \cite{drury1983introduction}, magnetic reconnection \cite{sironi2011acceleration}, and resonant absorption \cite{Hoshino:1992zz}.
DSA is one of the most widely used theories for high-energy particle acceleration, successfully explaining numerous high-energy phenomena \cite{klecker1998anomalous, stanev2010high}. It is frequently applied to astrophysical observations, such as solar physics, supernova remnants (SNR) and PWN \cite{hillas2005can, reynolds2008supernova,florinski2013galactic, kroon2016electron, amato2024particle}. Taking PWN as an example, relativistic electrons in pulsar winds are expelled outward until they encounter interstellar gas, and the wind decelerates through interactions. A shock interface forms between the relativistic pulsar wind and the decelerated wind, known as the termination shock (TS). Electrons transported across the shock into the downstream, then diffuse back upstream due to scattering in the downstream turbulent magnetic fields. Through repeated crossings, charged particles gain energy incrementally. The accelerated energy spectrum follows a power law, whose exponent depends on factors such as upstream/downstream velocities and diffusion coefficients.
Grounded in solar high-energy observations, this theory demonstrates robust validation within the MeV energy regime \cite{florinski2003galactic, kota2013theory}. While extrapolated to sub-PeV energy cosmic rays, its predictions remain unverified through direct experimental observation.
This work aims to test the DSA theory with the latest gamma-ray observations of the Geminga region above TeV energies.

The accelerated particles primarily exhibit diffusive behavior in cosmic propagation due to the interstellar turbulent magnetic fields. The typical value of the diffusion coefficient, found from fitting to cosmic ray data, is approximately $(3-5) \times 10^{28}  \text{cm}^2 \text{s}^{-1}$ for $\sim \text{GeV}$ particles, implying that the interstellar turbulent magnetic field strength is on the order of $\sim \mu \text{G}$ \cite{strong2007cosmic}. Moreover, various processes, such as cosmic ray streaming instability and SNR shock, are believed to have the potential to inject energy into the turbulent magnetic field and create localized slow-diffusion regions \cite{amato2011streaming, fang2019possible}. 
In the SNR induced model, the early SNR shock sweeps through the interstellar medium and excites the turbulence magnetic field with specific wave vectors. These turbulence evolves and propagates over time, eventually leading to a nearly uniform diffusion coefficient throughout the PWN and SNR. The turbulent magnetic fields significantly suppress the diffusion of lower-energy electrons, resulting in a reduced diffusion coefficient at low energies. However, as the electron energy approaches the sub-PeV, the diffusion coefficient increases rapidly until it becomes consistent with the typical interstellar value.

The multi-band energy spectra and angular distribution measurements of the Geminga halo, particularly the precise energy spectrum observations by HAWC \cite{albert2024precise} and LHAASO-KM2A\cite{guo2021observations} in recent years, have provided a rare opportunity to test these theories. The Geminga region, located at a distance of $250^{+230}_{-80}$ pc from Earth, is identified as a PWN with extended gamma-ray emission \cite{manchester2005australia}. 
The extended TeV emission spans $\sim 10^{\circ}$ around Geminga pulsar \cite{abdo2007tev,abeysekara2017extended}, and complementary observations by Fermi-LAT have revealed an extensive gamma-ray region around Geminga in the GeV range \cite{di2019detection}.
Observations above TeV energies show no significant difference in gamma-ray intensity between the inside and outside of the Geminga PWN, which is consistent with the expectations of the SNR-induced model.
Analysis of observational data suggests that the diffusion coefficient within the halo is approximately two orders of magnitude lower than the average diffusion coefficient in the general interstellar medium. Several theoretical models have been proposed to explain the pulsar halo, sharing a common framework: there is an electron slow-diffusion zone extending tens of parsecs around the pulsar \cite{liu2019understanding, evoli2018self, fang2019possible}. Within this region, high-energy electrons ($>\text{TeV}$) accelerated by the Geminga PWN remain confined and interact with ambient photon fields through inverse Compton (IC) scattering, generating the observed large-scale gamma-ray emission.

X-ray observations by experiments such as XMM-Newton and Chandra \cite{caraveo2003geminga,pavlov2010new,hui2017rapid} have revealed the PWN associated with Geminga, and provided more information about the upstream of the PWN termination shock. The Geminga PWN displays rich morphological features, including three distinct tail-like structures behind the proper motion of pulsar. These features are thought to result either from interactions between the PWN and the interstellar medium or from polar outflows originating within the PWN itself \cite{Posselt_2017}. 
The radius of the TS near the pulsar is thought to be $\sim 0.01$ pc \cite{pellizzoni2011detection}, which may be related to the age of the Geminga pulsar \cite{giacinti2020halo}. 
For particle acceleration in the PWN, this implies that the upstream of TS may be exceptionally small. 
Furthermore, the flux of X-ray radiation is relatively weak. The X-ray radiation observed by the Chandra experiment at keV energies is approximately $10^{32}$ erg/s, whereas the radiation flux in the Geminga region observed by Fermi-LAT and HAWC beyond GeV energies is about $5\times 10^{33}$ erg/s. This indicates that only a small fraction of the energy is lost through synchrotron radiation.During the acceleration and propagation of electrons, energy loss due to synchrotron radiation can be considered negligible.
Contrary to PWN observations, these experiments failed to detect the X-ray signal from the Geminga halo, possibly due to its relatively weak magnetic field. According to the calculation \cite{manconi2024geminga}, the average magnetic field in the Geminga halo could be less than $3 \mu G$.

This study proposes a validation method that tests the validity of both the DSA and the diffusion model with the spectral and profile measurements of the Geminga halo. In the DSA model, the diffusion coefficient influences the accelerated electron spectrum, which in turn affects the gamma-ray spectrum of the Geminga halo. In the diffusion model, the diffusion coefficient also influences the morphological profile of the halo. If the electron diffusion coefficient can simultaneously explain both the spectrum and the morphology, it provides strong evidence supporting the validity of both the DSA and diffusion models. The specific steps of this validation method are as follows:(1) Investigate the relationship between the diffusion coefficient and the electron spectrum or the halo morphology. (2) Derive the relationship between the upstream and downstream diffusion coefficients using X-ray and gamma-ray observations. (3) Fit the diffusion coefficient and other parameters based on halo observations. (4)Evaluate whether the fitting results are self-consistent.

This paper will proceed in the following sequence: Section \ref{sec2} will detail the specific procedures for deriving accelerated electron spectrum, and the relationship between the diffusion coefficient in the PWN and the halo. Section \ref {sec3} details the model fitting to  the energy spectrum of the Geminga halo. The diffusion coefficient derived from the fitting is found to be consistent with that derived from the halo profile. Section \ref{sec4} gives a summary of this work.

\section{Methods}\label{sec2}

To validate the DSA and diffusion models, we calculate the gamma-ray emission predicted for the Geminga halo.

The calculation of gamma-ray emission proceeds in three steps. (1) We model electron acceleration at the TS and derive the relation between the accelerated electron spectrum \(f_0(p)\) in the PWN and the upstream diffusion coefficient \(D^P\). (2) From $f_0(p)$ we derive the electron injection spectrum into the halo, $Q(E)$, and, together with the halo diffusion coefficient $D^{\mathrm{H}}$, compute the electron density in the halo, $N(E,\vec{r},t)$. (3) Using $N(E,\vec{r},t)$, we calculate the gamma-ray spectral and morphological distributions, $F_g(E_g,\vec{r})$. This workflow allows us to test the mutual consistency between the spectral and morphological constraints.


The high-energy electrons in the Geminga region gain energy through DSA on the TS, where their diffusion behavior plays a decisive role in shaping the accelerated electron spectrum. When these accelerated electrons diffuse far downstream of the TS, their energy loss becomes non-negligible. These high-energy electrons produce synchrotron radiation in ambient magnetic fields and generate gamma-ray emission through IC scattering with both the Cosmic Microwave Background (CMB) and infrared radiation fields from interstellar dust, which can be used to constrain the model parameters.

This section also discusses the diffusion coefficients in the PWN (upstream) and in the halo (downstream). The electron density is determined by the diffusion coefficient of the region. By calculating the electron density required to produce the X-rays and the electron density needed in the halo center for the gamma rays, the relationship between the diffusion coefficient in the PWN and the halo can be determined.

\subsection{electron acceleration on the TS}\label{sec21}

Assuming an isotropic particle density in phase space, DSA \cite{drury1983introduction} can be described in a simple form through the equation \ref{equ:acc}:

\begin{equation}\label{equ:acc} 
\begin{split}
\frac{\partial f(p,\vec{r},t)}{\partial t} +\vec{u}\cdot \nabla f(p,\vec{r},t) = \nabla (D(\vec{r},p)\nabla f(p,\vec{r},t)) \\
+\frac{p}{3}\nabla\cdot\vec{u}\frac{\partial f(p,\vec{r},t)}{\partial p} + q_{_{0}}
\end{split}
\end{equation}
where \( f(p,\vec{r},t) \) represent the phase space density of electron in the acceleration region, \( \vec{u} \) is the abbreviation of $u(\vec{r})$ denoting the velocity of gas in the upstream and downstream of the shock, \( D(\vec{r},p) \) represents the diffusion coefficient of electrons, $q_0$ stands for $q_{0}(p_{0})\delta(\vec{r} - \vec{r_{s}})$, which is the injected electrons with momentum $p_0$ at the TS and \( p \) is the electron momentum. 

As introduced in section \ref{sec1}, Geminga is an old pulsar whose PWN exhibits weak synchrotron radiation. As a result, energy loss of electrons within the nebula is relatively small, and thus the synchrotron energy loss term is neglected in equation \ref{equ:acc}.
Based on observations from Chandra and other experiments of the Geminga PWN, it is inferred that the magnetic field is approximately \( \sim 20\,\mu\mathrm{G} \) \cite{Posselt_2017}. Under such a magnetic field strength, the Larmor radius of accelerated electrons \( R_{L} \) is smaller than the TS radius \( R_{s} \simeq 0.01\,\mathrm{pc} \) \cite{giacinti2020halo} (\( R_{L} < R_{s} \)). Consequently, equation \ref{equ:acc} can be simplified to a a planar shock equation: \begin{equation}\label{equ:1d}
\begin{split}
u(x)\frac{\partial f}{\partial x} = \frac{\partial}{\partial x}\left(D(x,p)\frac{\partial f}{\partial x}\right)+\frac{p}{3}\frac{\partial u(x)}{\partial x} \frac{\partial f}{\partial p} \\
+ q_{_{0}}(p_{_{0}})\delta(x)
\end{split}
\end{equation}
where \( u(x) \) represents the gas velocities in upstream and downstream of the shock front, we define the position of shock front as \( x=0 \). The velocity profile is given by: 
\[ u = \begin{cases} u_1, & x < 0 \\ u_2, & x \geq 0 \end{cases} \] 
where \( u_1 \simeq c\) denotes the upstream velocity of the Geminga TS and \( u_2 \) denotes the downstream velocity. If boundary conditions \( f(- \infty) = 0 \) are imposed on this 1-D equation, the solution would be a power-law spectrum \cite{drury1983introduction}, which follows \( f_{0}(p)\propto p^{-\sigma}\). Where the $\sigma \equiv \frac{3u_1}{u_1-u_2}$, which characterizes the spectral slope of accelerated electrons. 

However, in practical scenarios where the upstream extent of the shock is finite, such boundary conditions lead to distinct electron spectra. In fact, there exists more than one condition leading to the finite upstream scale. One scenario involves electrons being captured by the pulsar at the far upstream (assumed to be at position \( x = -a \)). In this case, the shock upstream boundary condition can be written as \( f|_{x=-a} = 0 \), which is called zero-particle condition. The solution to the equation can then be readily obtained \cite{caprioli2009escape}:
\begin{equation}\label{equ:zp1}
    f(x,p) =
\left\{\begin{matrix}
\begin{split}
   f^{p}_{0}(p) \frac{\exp\left[\frac{u_1 x}{D^{P}(p)}\right] - \exp\left[-\frac{u_1 a}{D^{P}(p)}\right]}{1 - \exp\left[-\frac{u_1 a}{D^{P}(p)}\right]} ,\\
    x \le 0 
\end{split}\\
f^{p}_{0}(p), \quad  x \ge 0
\end{matrix}\right.
\end{equation}

where the accelerated electron spectrum \( f^{p}_{0}(p) \) at the shock surface is given by:
\begin{equation}\label{equ:zp}
    f^{p}_{0}(p) = K \exp\left\{-\sigma \int_{p_0}^{p} \frac{d\ln p'}{1 - \exp\left[-\frac{u_1 a}{D^{P}(p')}\right]} \right\}
\end{equation}

$D^{P}(p)$ is the diffusion coefficient in the PWN, or the diffusion coefficient in the upstream. And like most of the literature, it is simplified and does not change with position. \( K \) is a normalization constant and \( p_0 \) represents the initial momentum of injected electrons.

Another scenario occurs when electrons are reflected and cooled in the far upstream, preventing them from propagating further upstream. This condition bears strong analogy to the TS of solar wind, and the detailed process can be found in reference \cite{florinski2003cosmic}. The upstream boundary condition at the shock front adopts a zero-streaming condition: 
\begin{equation} 
D^{P}(p)\frac{\partial f}{\partial x}\bigg|_{x=-a}+\frac{u_1}{3}\frac{\partial f}{\partial \ln p}\bigg|_{x=-a}=0 
\end{equation}

The accelerated electron energy spectrum at the shock surface, \( f^{s}_{0}(p) \), is then given by: 
\begin{equation} \label{equ:zs}
f^{s}_{0}(p)=K \frac{3u_1}{u_1-u_2}\int^{\infty}_{p}f_{1}(p')d\ln p' 
\end{equation} 
where 
\begin{equation} \label{equ:zs1}
\begin{split}
    \frac{d\ln f_1}{d\ln p}=&-\sigma-\frac{\sigma-3}{e^{\eta}-1}+\\
    &\frac{\partial \ln D^{P}(p)}{\partial \ln p}\frac{ \eta}{e^{\eta }-1}
\end{split}
\end{equation}
where $\eta = \frac{u_1 a}{D^{P}(p)}$. For detailed derivations of these two conditions, please refer to Appendix \ref{ap1}.

Notably, when the diffusion coefficient is small ($D^{\mathrm{P}}(p) \ll u_{1} a$), $f_{0}(p)$ becomes a power law, which come back the classic first-order Fermi acceleration result. This arises because electrons with small diffusion coefficients have diffusion lengths \( \lambda_{d}\ll a \), making their acceleration and propagation not affected by the finite upstream. Moreover, the energy loss of electrons is negligible, leading to a power-law energy spectrum without the spectral break typically observed in pulsars around a few hundred GeV. Since this study focuses on electrons above 1 TeV, a single power-law offers an adequate and accurate description of the spectrum in this energy range.

On the other hand, both scenarios demonstrate that the electron energy spectrum depends on the diffusion coefficient. For high-energy electrons \( \lambda_{d} \ge a \), an exponential cut-off spectrum is given in both cases, and the shape of the cut-off is affected by the diffusion coefficient, which means that information about the diffusion coefficient is also contained through the energy spectrum.

\subsection{electron propagation through the Geminga halo}

The injection spectrum of halo is determined by the electron probability density \( f_{0}(p) \) from the previous subsection, and it can be written as
\begin{equation} 
Q(E,\mathbf{r},t) = \begin{cases} \frac{4\pi}{c} p^{2}f_{0}(p)\delta(\mathbf{r}-\mathbf{r_{s}})\frac{(1 + t/t_{sd})^{-2}}{(1 + t_s/t_{sd})^{-2}}, & t \ge 0 \\
0, & t < 0 \end{cases} 
\end{equation}

\( Q(E,\mathbf{r},t) \) is the source injection power. Here, \( \mathbf{r_{s}} \) is the position of pulsar. The spin-down characteristic timescale of Geminga is \( t_{sd} = 10 \, \text{kyr} \), and the pulsar age is taken as \( t_{s} = 342 \, \text{kyr} \). The conversion efficiency \( \epsilon \), representing the fraction of the  total energy loss of pulsar converted into high-energy electrons above 1 GeV, satisfies:

\begin{equation} 
\int^{\infty}_{1  \text{GeV/c}} 4\pi c p^{3}f_{0}(p)dp = \epsilon L \end{equation}

where $L$ is the current energy loss power. For the Geminga pulsar, \( L = 3.26 \times 10^{34} \, \text{erg/s} \).

The propagation of electrons in the halo is described by the following equation:

\begin{equation}\label{equ1} 
\begin{split} 
\frac{\partial N(E,\mathbf{r},t)}{\partial t} = & \nabla(D^{H}(E)\nabla N(E,\mathbf{r},t)) \ +\\
& \frac{\partial(b(E)N(E,\mathbf{r},t))}{\partial E} + Q(E,\mathbf{r},t) \end{split} 
\end{equation}

This equation accounts for electron injection, diffusion, and energy loss, where \( E^{2} = p^{2}c^{2} + m_{e}^{2}c^{4} \) represents the electron energy, \( N(E,\mathbf{r},t) \) is the particle number density, \( D^{H}(E) \) denotes the diffusion coefficient in the halo and \( b(E) = -dE/dt \) represents the momentum loss rate of electrons.

\( Q(E,\mathbf{r},t) \) is the source injection power. We model Geminga as a point source with an accelerated electron spectrum \( q(E) \):

The energy loss coefficient \( b(E) \) in the equation is calculated according to the method given in literature \cite{fang2021klein}, where \( b = b_0(E) E^2 \). In summary, the current electron density distribution \( N(E,\mathbf{r},t=t_{s}) \) can be obtained by solving equation \ref{equ1}, specifically:

\begin{equation}\label{equ:solve} 
\begin{split}
N(E,\mathbf{r},t) = \int^{t}_{0}dt_{0}\frac{b(E_{s}(t_0))}{b(E)}\frac{1}{(\pi\lambda^{2}(t_{0},t,E))^{\frac{3}{2}}} \times \\
\exp\left(-\frac{|\mathbf{r}-\mathbf{r_{s}}|^{2}}{\lambda(t_{0},t,E)^{2}}\right)Q(E_{s}(t_{0})) 
\end{split}
\end{equation}

In this equation, \( E_{s}(t_{0}) \simeq E/[1 - b_{0}E(t - t_{0})] \) represents the energy of electrons injected into the halo. The characteristic diffusion length of electrons is given by \( \lambda^{2}(t_{0},t,E) = 4\int^{E_{s}(t_{0})}_{E}dE'\frac{D^{H}(E')}{b(E')} \), where \( \mathbf{r_{s}} \) denotes the position of the electron source.

\subsection{radiation of the electrons}

Very high-energy gamma photons in pulsar halo originate from inverse Compton scattering of electrons. The surface brightness of gamma photons, denoted as \(  F_{g}(E_{g},\theta,\phi) \), is calculated using the formula \cite{blumenthal1970bremsstrahlung}:
\begin{equation}\label{equ:2}
    \frac{d F_{g}(E_{g},\theta,\phi)}{d\theta\,d\phi} = \int_{E_{\text{min}}}^{E_{\text{max}}}\int_{0}^{l_{\text{max}}}  P_{e \to \gamma}(E, E_{g})N(E, \mathbf{r},t=0) dE\, dl  
\end{equation}

where \( N(E, \mathbf{r}, t=0) \) represents the current electron volume density, \( l \) is the distance to the gamma-ray source along the line of sight. \(l_{max}\) is the maximum length between earth and the gamma-ray source in calculation, in this work, $l_{max}\equiv 1\,kpc$. The upper limit of the integral of the electron energy $E$ is $E_{max}=1\text{PeV}$, while the lower limit of the integral is $E_{min}=1\text{GeV}$. \( E_{g} \) denotes the energy of gamma photons and $(\theta, \phi)$ denote the angular coordinates used to describe the solid angle of the observation region. The power density distribution for IC scattering \( P_{e \to \gamma}(E, E_{g}) \) is expressed as \cite{blumenthal1970bremsstrahlung}:
\begin{equation}
\begin{split}
    P_{e \to \gamma}(E, E_{g}) = &\frac{3c\sigma_T}{4\gamma_e^2} \int_{1/(4\gamma_e)}^{1} dk \, \frac{dn}{d\epsilon}(\epsilon(k)) \times \\
    &\left(1 - \frac{1}{4k\gamma_e(1 - E_{g}/E)}\right) \times \\
    &\left(2k\ln k + k + 1 - 2k^2 + \frac{E_{g}(1 - k)}{2(E - E_{g})}\right)
\end{split} 
\end{equation}

with the dimensionless parameters:

\[ \gamma_e = \frac{E}{m_e c^2}, \quad k = \frac{E_{\gamma} m_e^2 c^4}{4\epsilon E(E - E_{\gamma})} \]

Here, \( \sigma_T \) is the Thomson scattering cross-section, \( \epsilon \) denotes the energy of seed photons, and \( \frac{dn}{d\epsilon}(\epsilon(k)) \) represents the energy spectrum of the seed photon field. This study considers two types of seed photon fields: the CMB (\( T_{\text{CMB}} = 2.73 \, \text{K} \)) and infrared radiation from interstellar dust (\( T_d = 20 \, \text{K} \)). Detailed calculation methods can be found in reference \cite{blumenthal1970bremsstrahlung}.

\subsection{\label{sec24}The Diffusion coefficient in PWN and Halo}

\begin{figure}
    \centering
    \includegraphics[width=\linewidth]{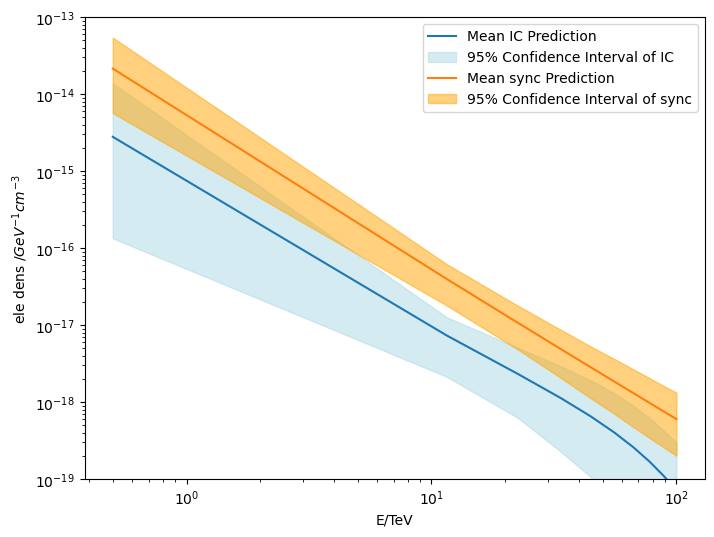}
    \caption{Electron density of the center derived from the measurement of Chandra (Orange Line) and HAWC (Blue Line). The orange region and blue region represents the 95\% confidence interval inferred from Chandra and HAWC measurements. The unit of y-axis has been converted to $GeV^{-1} cm^{-3}$, with the radius of Geminga PWN $r_{PWN}=0.01pc$ and the magnetic field $B_{PWN}=20\mu G$. The diffusion coefficient in the Geminga halo is set to be $D^{H}=4.5\times10^{27}(E/100TeV)^{1/3}$}
    \label{fig:diff1}
\end{figure}

Although the DSA theory suggests that the upstream and downstream diffusion coefficients are independent, in the case of old PWNs, the diffusion coefficient in the halo is quite similar to that in the PWN. This is because turbulent magnetic fields have propagated both inward and outward over a long period, leading to a more uniform diffusion coefficient throughout. To test this idea, the X-ray data from the PWN and the gamma ray data from the halo are involved in the comparison.

 Assuming spatial transport of electrons in the PWN is dominated by diffusion effect, the solution to the diffusion equation (equation \ref{equ:solve}) reveals a relationship between electron density and diffusion coefficient: \( N\propto D^{-3/2} \). Gamma-ray observations can determine the electron density in the halo (\( N^{H} \)), while X-ray observations constrain the electron density in the PWN (\( N^{P} \)). These measurements enable us to establish the diffusion coefficient ratio between the two regions:
\begin{equation}\label{equ:cp} 
\frac{D^{P}}{D^{H}} = \left(\frac{N^{H}}{N^{P}}\right)^{3/2} \end{equation}

Adopting a magnetic field of \( 20\,\mu\mathrm{G} \) and radius \( r_{\mathrm{PWN}}=0.01\,\mathrm{pc} \) for the PWN, we derived the electron density near 10 TeV through non-thermal radiation computational code Naima \cite{zabalza2015naima}, based on Chandra X-ray spectral measurements. The electron spectrum was modeled as an exponential cutoff power-law with a fixed cutoff energy of 100 TeV. The synchrotron fitting yields \( N_{e,x}=5.3^{+1.0}_{-1.9}\times10^{-17}\,\mathrm{GeV}^{-1} \mathrm{cm}^{-3} \) at 10 TeV, with a spectral index of \( 1.97^{+0.14}_{-0.19} \) (the orange line in figure \ref{fig:diff1}).

On the other hand, using equations \ref{equ:solve} and \ref{equ:2} with HAWC data points, we calculated the central electron density in the halo after setting the diffusion coefficient $D^{H}(E)=4.5\times10^{27}(E/100TeV)^{1/3} cm^{2}~s^{-1}$ . Modeling with an exponential cutoff power-law spectrum gives a central density of \( 9.8^{+3.6}_{-3.1}\times10^{-18}\,\mathrm{GeV}^{-1}\mathrm{cm}^{-3} \) at 10 TeV, spectral index \( 1.7^{+0.4}_{-0.2} \), and cutoff energy \( 88^{+10}_{-12}\,\mathrm{TeV} \) (the blue line in figure \ref{fig:diff1}). Figure \ref{fig:diff1} compares the electron number densities obtained through both methods, with shaded regions indicating respective 95\% confidence intervals.

Based on the results shown in figure \ref{fig:diff1}, we can draw two key conclusions. First, the power-law spectra of the two electron densities exhibit remarkable similarity, indicating that the diffusion coefficients in the PWN and halo share the same energy dependence. 

Second, The electron density required by synchrotron radiation is similar to the density required by IC radiation, although synchrotron radiation results need more electrons. In these calculations, we adopted $B_{PWN}=20\mu G$, a magnetic field strength estimated from Chandra X-ray observations of the tail region in the Geminga PWN. However, the actual magnetic field in the Geminga PWN could be significantly stronger. If the magnetic field in the Geminga PWN reaches $ B_{PWN}=44\mu G$, the diffusion coefficients in both the PWN and halo would become nearly the same. 

Taken together, these findings provide the justification for treating the diffusion coefficient $D(E)$ as identical in both the Geminga PWN and halo.
It should be noted that PWN can also generate turbulent magnetic fields that propagates outward. Considering an ambient Alfvén speed of $v_A \sim 30\,\mathrm{km/s}$, such turbulence can travel several parsecs over the spin-down timescale of Geminga. This turbulence would suppress diffusion in the surrounding region, implying that the diffusion coefficient inside the PWN is lower than in the halo, though not significantly. For simplicity, we assume the same diffusion coefficient for both the PWN and the halo in our calculations. We will later discuss the scenario in which the diffusion coefficient in the PWN is smaller than that in the halo, based on the fitting results presented in sections \ref{sec3} and \ref{sec4}.

\section{Result}\label{sec3}

\begin{figure}
    \centering
    \includegraphics[width=\linewidth]{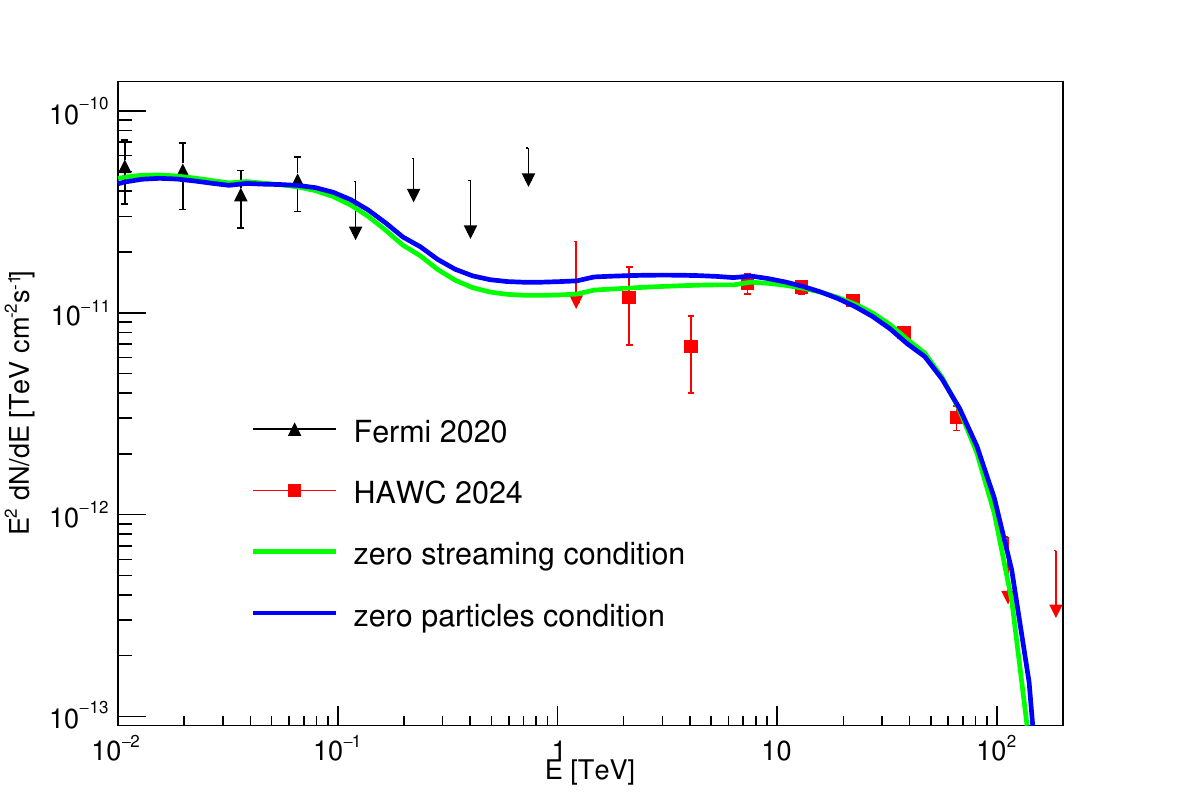}
    \caption{ Spectra fit of Geminga halo. The black data points are from Fermi-LAT and the red data points are from HAWC. Two different condition of electron acceleration in PWN with their best fit parameters and $\delta_1 =0.3$ are shown in figure. The blue line and green line shows the best fit zero-particle condition and zero-streaming condition respectively.   }
    \label{fig:twoc}
\end{figure}

\begin{figure*}
    \centering
    \subfigure[]{\includegraphics[width=0.46\linewidth]{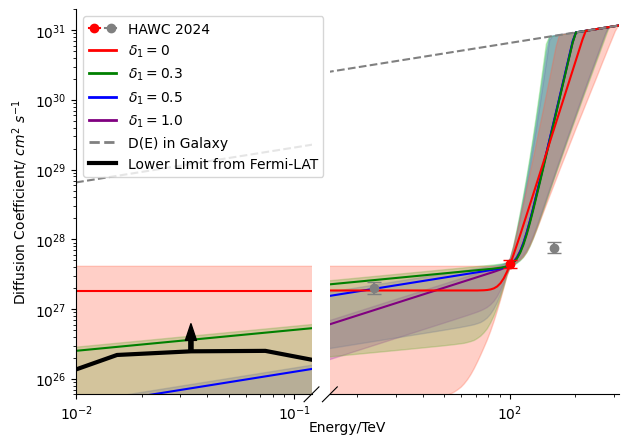}}
    \subfigure[]{\includegraphics[width=0.46\linewidth]{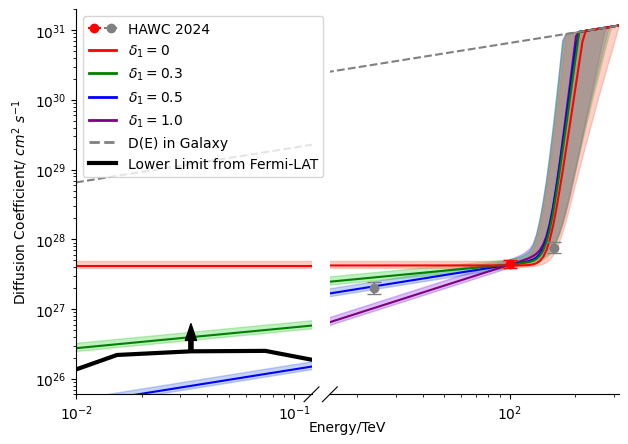}}
    \caption{ The diffusion coefficients corresponding to the fitted parameters under the condition that  $D_{100TeV} = 4.4^{+0.66}_{-0.57}\times 10^{27} cm^{2}~ s^{-1}$. The lines and shaded regions in different colors represent the best-fit diffusion coefficients and their corresponding 1-\(\sigma\) uncertainty ranges for different \( \delta_1 \) values. The gray dashed line indicates the average diffusion coefficient within the Galaxy. (a) the zero-streaming condition. (b) the zero-particle condition.}
    \label{fig:diff}
\end{figure*}

In the DSA model, the accelerated electron spectrum is influenced by the diffusion coefficient. When \( D > u_1 a \), the power-law of the diffusion coefficient $D(E)$ alters the shape of the electron spectrum. This implies that in this regime, the power-law index of $D(E)$ can be determined from the accelerated electron spectrum. However, at lower energies (\( D < u_1 a \)), the electron spectrum degenerates into a pure power-law $f_{0} \propto p^{-\sigma}$, becoming independent of the diffusion coefficient. Consequently, at low energies, $D(E)$ cannot be determined from the electron spectrum alone. Fortunately, observations of the morphology of the Geminga halo provide supplementary information about the diffusion coefficient at lower energies. The HAWC experiment, by observing the halo morphology at different energies, has inferred the diffusion coefficients corresponding to electrons with energies of 24 TeV, 100 TeV, and 160 TeV\cite{albert2024precise}. 
Meanwhile, an analysis of Fermi-LAT data indicated no gamma-ray emission from the Geminga halo within $10^{\circ}$ \cite{Xi:2018mii}, while another study indicated a detection extending to $\sim 30^{\circ}$ \cite{di2019detection}. Together, these results suggest that the diffusion coefficient cannot be too large—otherwise the emission would not be visible within $30^{\circ}$—nor too small, as this would result in detectable emission within $10^{\circ}$. Thus, Fermi-LAT observations implicitly constrain the diffusion coefficient at energies around $10\, \mathrm{GeV}$. 

According to the latest research of pulsar halo diffusion theory \cite{fang2019possible}, we adopt a break power-law for the diffusion coefficient to minimize systematic errors

\begin{equation}\label{equ:di}
   D(E) = \left\{\begin{matrix}
   &D_{c} (E/E_c)^{\delta_1} , E<E_c\\
    &D_{c}(E/E_c)^{\delta} , E\ge E_c
\end{matrix}\right.
\end{equation}
where \( D_c \) represents the diffusion coefficient at a specific energy \( E_c \), while \( E_c \) denotes the break energy of the diffusion coefficient. The parameters $\delta_1$ and $\delta$ correspond to the power-law index below and above the break energy \( E_c \), respectively.Under this diffusion coefficient $D(E)$, the model contains 7 unknown parameters: \( E_c\), \( u_1 a \), \(D_c\), \( \delta \), \( \delta_1 \), \(\sigma\) and the energy conversion efficiency \( \epsilon \). These parameters respectively govern the cutoff energy of the diffusion coefficient, the cutoff energy of electron spectrum, the post-cutoff power-law index, the pre-cutoff power-law index, and the spectral normalization. 

The gamma-ray spectra of the Geminga halo obtained by the Fermi-LAT and HAWC, along with the diffusion coefficients of electrons measured by the HAWC, are used to fit the parameters of these models. However, when fitting the electron spectrum, HAWC did not use the Fermi-LAT data, whereas this work performs a joint fit of both Fermi-LAT and HAWC data. The resulting electron spectra from the two fits exhibit different power-law indices, and the electron energy inferred from gamma-ray observations also differs. However, as the pivot energy of the HAWC experiment, the gamma-ray energy corresponding to 100 TeV electrons shows only a small difference under different electron spectral indices. Therefore, the measurement of the diffusion coefficient for 100 TeV electrons is reasonable for this model. On the other hand, we also attempt to fix the spectral index to the value obtained from the HAWC fit, and the fitted diffusion coefficient remain similar. Therefore, we use the spectra data measured by Fermi-LAT and HAWC in the Geminga region, along with the 100 TeV diffusion coefficient \(D_{100}\) measured by HAWC, to fit the model parameters. Due to the lack of low-energy diffusion coefficient information in the fitting data, we find it challenging to place constraints on \(\delta_1\). So, we systematically investigate its impact on parameter fitting by considering different values of \(\delta_1 = 0, 0.3, 0.5,\) and \(1.0\) as part of our uncertainty analysis.
In addition, there is a constraint that $D(E) \le D_{ISM}(E)\simeq 2.08\times10^{28}(E/GeV)^{0.5}cm^{2}~s^{-1}$, which means that $D(E)$ in halo should be less than the averange diffusion coefficient in the Galaxy \cite{yuan2017propagation}.

The likelihood function of the fitting is defined as:
\begin{equation}
    \mathcal{L}(\Theta) \propto \exp\left(-\frac{1}{2}\chi^{2}(\Theta)\right)
\end{equation}
where the \( \chi^{2}(\Theta) \) statistic is given by: 
\begin{equation}
\begin{matrix}
    \chi^{2}(\Theta)  =\sum^{N_{\text{data}}}_{i}&  \left(\frac{F^{\text{DSA}}_{i}(\Theta) - F^{\text{exp}}_i}{\sigma^{\text{exp}}_{i}}\right)^{2}+ \\
    &\left(\frac{D(100TeV)-D_{100}}{\sigma_{D}}\right)^{2}
    \end{matrix}
\end{equation}
Here, \( F^{\text{DSA}}(E_{i}) \) represents the gamma-ray flux of Geminga at energy \( E_i \), predicted by the DSA model under parameter set \( \Theta \), while \( F^{\text{exp}}_i \) and \( \sigma^{\text{exp}}_{i} \) denote the experimentally measured gamma-ray flux and its uncertainty for the entire halo. \( D_{100} \) and \( \sigma_D \) represent the diffusion coefficient and its uncertainty at 100 TeV as measured by HAWC, while \( D(100TeV) \) denotes the diffusion coefficient defined in equation \ref{equ:di} at 100 TeV. 
Parameter estimation employs Markov Chain Monte Carlo (MCMC) sampling methods implemented through the Python package emcee \cite{Foreman-Mackey_2013}.

\begin{table*}[]
\renewcommand{\arraystretch}{1.5}
    \centering
    \caption{Fitting results of parameters in the zero-streaming condition and the best $\chi^2$/n.d.f. with its $\epsilon$ at this time.}
    \begin{ruledtabular}
    \begin{tabular}{cccccccc}
       &$E_c$/TeV & $lg[u_1a/ (cm^{2}~s^{-1})]$ &  \( \delta \) & \( \sigma  \)  & $lg[D_c/ (cm^{2}~s^{-1})]$ & $\epsilon$ & $\chi^{2}$/n.d.f.\\
        \hline
       $\delta_1=0$& $92.9_{-26.1}^{+25.4}$ & $28.2_{-0.7}^{+0.9}$ &$9.8_{-2.3}^{+6.5}$ &$3.4_{-0.3}^{+0.3}$ 
& $27.6_{-0.6}^{+0.6}$ & 0.098 & 5.2/(12-6)\\
       $\delta_1=0.3$  & $112.1_{-29.3}^{+8.5}$ & $28.0_{-0.5}^{+0.8}$ &  $13.4_{-5.1}^{+11.3}$ &$3.5_{-0.2}^{+0.2}$ & $27.9_{-0.4}^{+0.5}$ & 0.045 & 5.3/(12-6)\\
       $\delta_1=0.5$  & $111.7_{-24.8}^{+9.0}$ & $28.1_{-0.5}^{+0.7}$ &  $13.3_{-5.0}^{+10.5}$ &$3.5_{-0.2}^{+0.2}$ & $27.9_{-0.4}^{+0.5}$ & 0.044 & 4.8/(12-6)\\
       $\delta_1=1.0$  & $112.1_{-21.4}^{+9.0}$ & $28.1_{-0.5}^{+0.6}$ &  $13.2_{-4.9}^{+9.4}$ &$3.4_{-0.2}^{+0.3}$ & $27.9_{-0.4}^{+0.5}$ & 0.034 & 5.5/(12-6)\\
    \end{tabular}
    \end{ruledtabular}
    \label{tab:1}
\end{table*}

\begin{table*}[]
\renewcommand{\arraystretch}{1.5}
    \centering
    \caption{Fitting results of parameters in the zero-particle condition and the best $\chi^2$/n.d.f. with its $\epsilon$ at this time.}
    \begin{ruledtabular}
    \begin{tabular}{cccccccc}
       &$E_c$/TeV & $lg[u_1a/ (cm^{2}~s^{-1})]$ &  \( \delta \) & \( \sigma  \) & $lg[D_c/ (cm^{2}~s^{-1})]$ & $\epsilon$ & $\chi^{2}$/n.d.f. \\
        \hline
      $\delta_1=0$& $151.1_{-19.5}^{+13.1}$ & $29.8_{-1.4}^{+1.2}$ &$20.7_{-10.1}^{+11.8}$ &$3.6_{-0.1}^{+0.1}$  & $28.3_{-0.6}^{+0.8}$ & 0.058 & 15.9/(12-6)\\
       $\delta_1=0.3$  & $149.9_{-18.1}^{+14.1}$ & $29.8_{-1.4}^{+1.2}$ &  $22.9_{-11.3}^{+11.2}$ &$3.4_{-0.1}^{+0.1}$& $28.4_{-0.7}^{+0.8}$ & 0.044 & 13.3/(12-6)\\
       $\delta_1=0.5$  & $149.4_{-17.7}^{+15.1}$ & $29.7_{-1.5}^{+1.3}$ &  $23.5_{-11.5}^{+10.6}$ &$3.3_{-0.2}^{+0.1}$& $28.4_{-0.7}^{+0.9}$ & 0.034 & 11.8/(12-6)\\
       $\delta_1=1.0$ & $151.5_{-17.9}^{+19.5}$ & $29.6_{-1.5}^{+1.3}$  &  $25.0_{-12.0}^{+9.9}$ &$3.1_{-0.1}^{+0.2}$& $28.5_{-0.8}^{+1.0}$ & 0.028 & 11.9/(12-6)\\
    \end{tabular}
    \end{ruledtabular}
    \label{tab:2}
\end{table*}

The two electron acceleration spectra mentioned in section \ref{sec21} are respectively fitted to the observations. Since the gamma-ray spectra are identical for different \( \delta_1 \) values, each of the two conditions is represented by a single curve with \( \delta_1 = 0.3 \). The gamma-ray spectra corresponding to the best-fit parameters are shown in figure \ref{fig:twoc}. The blue line represents the gamma-ray spectrum under the zero-particle condition for the Geminga halo, while the green line corresponds to the spectrum under the zero-streaming condition. The spectra predicted by the model shows excellent agreement with observations. The best-fit parameter values and their uncertainties are listed in table \ref{tab:1} and \ref{tab:2}. Since the two conditions have different electron spectral shapes, their fitting results exhibit significant discrepancies. By looking at the $\chi^2$/n.d.f., the zero-streaming condition produces a smaller chi-square than the zero-particle condition, indicating better consistency with observations.

The HAWC experiment, through morphological observations of the Geminga Halo, estimated the diffusion coefficient for 100 TeV electrons to be $D_{100} = 4.4^{+0.66}_{-0.57}\times 10^{27} cm^{2}~ s^{-1}$. By putting the fitted diffusion coefficient parameters into equation \ref{equ:di}, we obtain diffusion coefficient ranges under two different conditions, as shown in figure \ref{fig:diff}. In the figure, the lines and shaded regions in different colors represent the best-fit diffusion coefficients and their corresponding 1-\(\sigma\) uncertainty ranges for different \( \delta_1 \) values. The figure \ref{fig:diff}(a) shows the diffusion coefficient under the zero-streaming condition, while the figure \ref{fig:diff}(b) represents the diffusion coefficient under the zero-particle condition. The gray dashed line indicates the average diffusion coefficient within the Galaxy. We incorporate the Fermi-LAT upper limit on the flux within $10^{\circ}$ around Geminga into the model calculation, and the resulting lower bound on the diffusion coefficient is shown as a solid black line in the figure \ref{fig:diff}.

The results indicate that, for any \( \delta_1 \) value, the diffusion coefficient \( D(E) \) must exhibit a rapid increase around 100 TeV, and approach the Galactic value at higher energies ($\sim$ 200 TeV). In addition, the Fermi-LAT upper limit on the gamma-ray flux within $10^{\circ}$ of the Geminga halo requires the diffusion coefficient in the halo to be greater than $10^{26}~\mathrm{cm}^2\,\mathrm{s}^{-1}$ at $\sim$ 10 GeV, which is consistent with the value estimated by the Fermi-LAT analysis \citep{di2019detection}. As shown in figure \ref{fig:diff}, the fitted results of $\delta_1 = 0$ and $\delta_1 = 0.3$ for both conditions are in good agreement with the lower bound on the diffusion coefficient inferred from the Fermi-LAT $10^{\circ}$ observation.

The parameter $u_1 a$ determines the cutoff energy of the electron spectrum, which means that a larger value corresponds to a higher cutoff energy. In the model, the shock upstream consists of a relativistic pulsar wind with a velocity approaching the speed of light (\( u_1 \sim c \)). Through X-ray observations, we estimate the size of the TS to be approximately 0.015 pc, which provides a strict constraint \( \log (u_{1}a) \sim 27.2 \). Under this limitation, the fitting results for each $\delta_1$ condition exhibit a significance within $2-\sigma$.
According to Equations \ref{equ:zp1} and \ref{equ:zs1}, the diffusion coefficient $D^p$ and the model parameter $u_{1}a$ are degenerate. The fitted values of $u_1 a$ are generally larger than the observed values, which may indicate that the diffusion coefficient inside the PWN is smaller than that in the halo. If the PWN diffusion coefficient is assumed to be 1/5 of that in the halo, then under both boundary conditions, the fit values of $u_1 a$ are in good agreement with experimental observations.

For the parameter $\sigma$, the zero-streaming condition gives $ 3.4 - 3.5$ while the zero-particle condition yields $3.1 - 3.6$. In ultrarelativistic scenarios, the velocity relationship $u_1 = 7u_2$ corresponds to $\sigma \ge 3.5$ \cite{kirk2000particle}. Under the zero-streaming condition, each $\delta_1$ fit well within the \( \sigma \) constraints, while under the zero-particle condition, \( \delta_1 = 0 \) and \( \delta_1=0.3\) provides a good fit to this constraint.

The index of the diffusion coefficient $\delta$ deviates significantly from conventional Kolmogorov diffusion.
Taking $\delta_1$ = 0.3 as an example, the best-fit values of the diffusion index $\delta$ under the two conditions are 13.4\(^{+11.3}_{-5.1}\) and 22.9\(^{+11.2}_{-11.3}\), respectively. All these $\delta$ values differ significantly from the Kolmogorov diffusion, making it challenging to describe the results within the traditional diffusion models. But under the SNR induced model, this phenomenon can be explained naturally, which means that Geminga SNR cannot excite a turbulent magnetic field with wave vector $k<1/r_{L}$, where $r_L$ represent the Larmor radius of cutoff energy in accelerated electron spectum.

Although our fitting only adopts the 100 TeV diffusion coefficient measured by HAWC, the diffusion coefficients at 24 TeV and 160 TeV can still be used to assess the validity of model. 
As shown in figure \ref{fig:diff}, under the zero-streaming condition, the deviations of $\delta_1$ = 0, 0.3 and 0.5 from the 24 TeV point measured by HAWC have a significance within $1-\sigma$ range, while their deviations from the HAWC 160 TeV measurement are $2.4 \sigma$, $2.3 \sigma$ and $2.3 \sigma$, respectively. Under the zero-particle condition, $\delta_1$ = 0.3 and 0.5 exhibits good agreement with the HAWC measurement points, whereas $\delta_1= 0$ and 1.0 shows a deviation with a significance of $3.0 \sigma$ from the 24 TeV diffusion coefficient. Overall, the observations of Geminga halo tend to favor the cases with $\delta_1 = 0.3$ and $\delta_1 = 0.5$.

\begin{figure*}
    \centering
    \includegraphics[width=\linewidth]{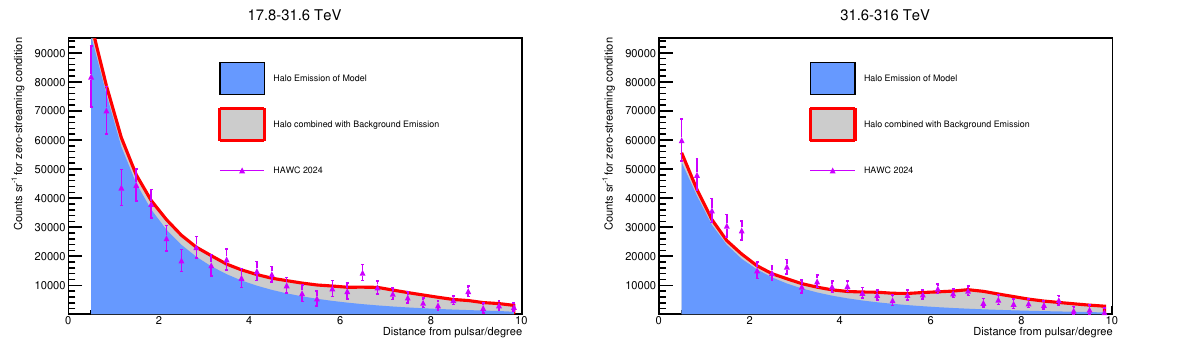}
    \includegraphics[width=\linewidth]{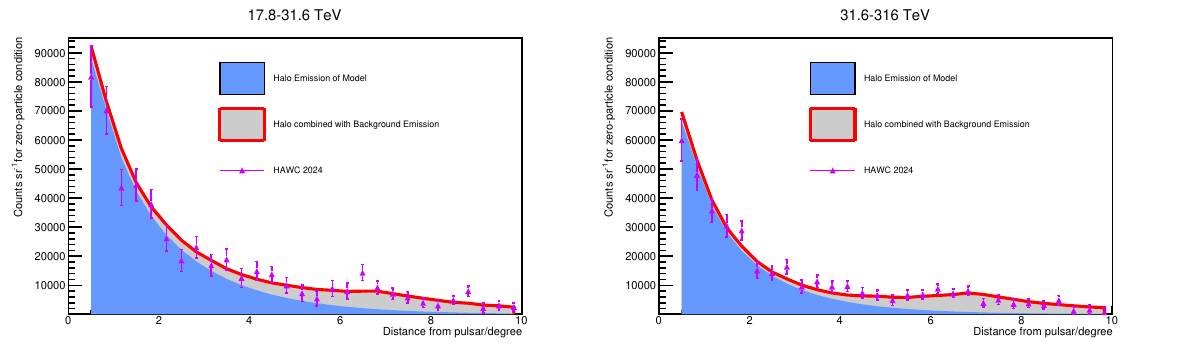}
    \caption{Brightness profile of Geminga halo measued by HAWC in two energy bins. The top two figures are profiles under zero-streaming condition, and the bottom two figures are profiles under zero-particle condition. The first energy bin is 17.8-31.6 TeV and the second energy bin is 31.6-316 TeV. Blue region is the gamma-ray profile predicted by this model. Combining with the background gamma-ray emission given by HAWC, the total profile is shown in red line, which describes the gamma-ray profile well. 
    }
    \label{fig:morph}
\end{figure*}

HAWC also released their observations of the Geminga halo profile across three energy bands, whose energy are 1 TeV- 17.6 TeV, 17.6 TeV- 31.6 TeV and 31.6 TeV- 316 TeV. Since the authors only found that the effective area of HAWC becomes stable above 10 TeV, only the last two energy ranges are studied and an identical effective area (40000 $m^{2}$) is adopted in the analysis. Figure \ref{fig:morph} presents the Geminga halo morphology for a specific set of parameters (in the zero-streaming condition, \( \delta = 8 \), \(\delta_1=0.5\), \( \sigma = 3.3 \), and \( E_c = 100 \) TeV; in the zero-particle condition, \( \delta = 10 \), \(\delta_1=0.5\), \( \sigma = 3.3 \), and \( E_c = 150 \) TeV). The blue region in the figure represents the gamma-ray profile predicted by the model, while the gray area accounts for contributions from other extended sources, such as Monogem, to the Geminga observations. Under these parameter settings, the Geminga halo profiles of this model agree well with the observations, despite significant differences between the model-predicted diffusion coefficients and those provided by HAWC at 24 TeV and 160 TeV. This discrepancy indicates that this model differs from the one used in the HAWC analysis. Therefore, the diffusion coefficients at 24 TeV and 160 TeV given by HAWC could be considered as references rather than strict constraints for excluding model parameters.

\section{Summary and Discussion}\label{sec4}

This study proposes to test the validity of DSA model combined with a diffusion model called SNR induced model with the unique observations of Geminga PWN and halo. By applying the model to the Geminga halo, we comprehensively explain the gamma-ray spectral cut-off and its morphological observations. In this work, a bounded upstream region of the pulsar TS has been introduced and accelerated electron spectra under two different cut-off conditions are derived. We constrained parameters through joint fitting with Fermi-LAT and HAWC data, obtaining the diffusion coefficient of Geminga region and predicting a rapidly increasing after 100TeV. The best energy spectrum parameters are in good agreement with the profile and the theories of DSA and diffusion propagation are consistent with experimental observations. Finally, this work has not confirmed or excluded other models and similar tests would be both important and valuable for other models, such as magnetic reconnection acceleration \cite{sironi2011acceleration}.

In the calculations of section \ref{sec3}, we assumed that the PWN has the same diffusion coefficient as the halo. However, results in this work suggest that the diffusion coefficient in the PWN may be five times lower than in the halo, possibly due to the magnetic turbulence generated by the PWN itself. Combined with Chandra X-ray observations, this also implies a magnetic field strength of about 20 $\mu G$ in the Geminga PWN.

On the other hand, this model can be further improved. For instance, the width of the shock upstream \( a \) may depend on electron energy. If an energy-dependent \( a(E) \) is considered, the final cutoff shape of the electron spectrum will also be altered. Additionally, in solving the acceleration equation, we adopted a planar approximation. However, to obtain a more precise solution for the electron spectrum, solving under a spherical shock would be better. Uncertainties in the median energy of the diffusion coefficients measured by HAWC can strongly influence parameter fittings. If the median energy is lower than the actual value, this would lead to a smaller fitting \( u_1 a \). Investigating this uncertainty further could be helpful to reduce parameter errors.

Besides, the amount of data on other pulsar halo, like Monogem, have been lacking to help validate the model. Future experiments with enhanced sensitivity beyond the $100$ TeV energy range, such as LHAASO-KM2A, will provide more precise observation required to rigorously validate our works \cite{cao2019large}. Furthermore, our results indicate that a cutoff in the diffusion coefficient may be a universal feature of high-energy particle acceleration sources. This finding is consistent with the calculations in Ref.~\cite{Mukhopadhyay:2021dyh}, and suggests that self-generated cosmic-ray turbulence could be responsible for the formation of pulsar halos. For PeVatrons like Crab, which can produce PeV gamma rays, the cutoff energy of diffusion coefficient would be much higher than that of Geminga, possibly above 10 PeV. Therefore, more multi-wavelength observations will be able to perform better tests for the acceleration and propagation model.

\begin{acknowledgements}
    We thank XiaoJun Bi, Qiang Yuan, Yi Zhang, YingYing Guo and Dan Li for valuable discussions about this work. This work is supported by National Natural Science Foundation of China (NSFC) (Nos. 12333006, 12105292, 12393853). This work is supported by the Ministry of Science and Technology of China, National Key $R\&D$ Program of China (2018YFA0404203).
\end{acknowledgements}

\appendix

\section{Solution of the One-Dimensional Bounded Electron Acceleration Equation}\label{ap1}
Although the solution to equation \ref{equ:1d} with the boundary condition \( f(- \infty) = 0 \) takes a power-law form, the situation becomes more complex when the upstream region is bounded. Here, the boundary condition requires the electron density to vanish at \( x = -a \), i.e., \( f|_{x=-a} = 0 \), and the second term on the right-hand side of equation \ref{equ:1d} becomes zero because the gas velocities in the upstream and downstream regions are constant. Under these conditions, the equation reduces to

\begin{equation}\label{equ:a1}
     u\frac{\partial f}{\partial x}  = \frac{\partial}{\partial x}(D(p)\frac{\partial f}{\partial x})
\end{equation}

The general solution is given by
    \begin{equation}\label{equ:a2}
    f(x,p) = g_1(p)exp(\int^{x}_{0}\frac{u(x')}{D(x',p)}\text{d}x')+g_2(p)
\end{equation}
where $g_1(p)$ and $g_2(p)$ are arbitrary functions of momentum. To ensure that $f(x,p)$ remains finite at infinity, the solution must take the form
\begin{equation}
    f(x,p)=\left\{\begin{matrix}
 f_1(p)+g_1(p)exp(\frac{ux}{D(p)}) ,x<0\\
 f_2(p) , x\ge 0
\end{matrix}\right.
\end{equation}
with $f_1(p)$and $f_2(p)$ also being functions of momentum. Furthermore, at the shock surface $x=0$, both the particle distribution function and the particle flux must be continuous. These conditions imply

\begin{equation}
    f_1(p)+g_1(p)=f_2(p)
\end{equation}
and
\begin{equation}
    u_1p\frac{\partial}{\partial p}(f_1(p)+g_1(p))+3u_1g_1 = u_2 p \frac{\partial }{\partial p}f_2
\end{equation}
By combining the above equations and applying the boundary conditions—such as when the shock upstream momentum distribution $f_1(p)$ is known—the accelerated particle momentum distribution downstream can be expressed as:
\begin{equation}
    f_2(p) = \sigma p^{-\sigma}\int^{p}_{0}p'^{\sigma-1}f_1(p')\text{d}p'+c_0p^{-\sigma}
\end{equation}
where $c_0$ is a constant. Substituting the boundary condition $f|_{x=-a}=0$, the solution to equation \ref{equ:1d} can be derived as
\begin{equation}
    f(x,p)=f_{0}(p)\frac{exp[\frac{u_1x}{D(p)}]-exp[-\frac{u_1a}{D(p)}]}{1-exp[-\frac{u_1a}{D(p)}]} 
\end{equation}
where
\begin{equation}
    f_{0}(p)= f_0 exp\{-\sigma\int^{p}_{p_{0}}\frac{\text{dln}p'}{1-exp[-\frac{u_1a}{D(p')} ]}  \}
\end{equation}
Here, $f_0(p)$ can be interpreted as the momentum-dependent part of the electron distribution, and $f_0$ is a constant.

In an alternative scenario, electrons cannot propagate further upstream due to reflection and cooling in the far-upstream region. The corresponding upstream boundary condition at the shock then becomes a zero diffusion flux condition: 
\begin{equation} \label{equ:ab1}
D(p)\frac{\partial f}{\partial x}\bigg|_{x=-a}+\frac{u_1}{3}\frac{\partial f}{\partial \ln p}\bigg|_{x=-a}=0 
\end{equation}

Because the downstream distribution function is independent of \(x\), the streaming flux at the shock must vanish. Therefore,
\begin{equation}\label{equ:ab2}
D(p)\frac{\partial f}{\partial x}\bigg|_{x=0}+\frac{u_1}{\sigma}\frac{\partial f}{\partial \ln p}\bigg|_{x=0}=0
\end{equation}

Under this condition, the diffusion equation in both the upstream and downstream can still be written in the form of equation ~\ref{equ:a1}. For convenience, the general solution for \(f(x,p)\) in equation ~\ref{equ:a2} can be rewritten as
\begin{equation}\label{equ:ag1}
f(x,p)=\left\{\begin{matrix} f_1(p)(exp(\frac{ux}{D})-1+f_2(p)) ,x<0\\ f_1(p)f_2(p) , x\ge 0 \end{matrix}\right.
\end{equation}

In this case, substituting the general solution \ref{equ:ag1} directly into the boundary conditions \ref{equ:ab1} and \ref{equ:ab2} yields
\begin{equation}\label{equ:aj1}
\frac{\text{d ln} f_1}{\text{d ln} p} = -\sigma -\frac{\sigma -3}{exp{[\frac{u_1a}{D(p)}]} -1} + \frac{\alpha u_1a}{D(p)(exp[\frac{u_1a}{D(p)}]-1)}
\end{equation}
and
\begin{equation}\label{equ:aj2}
\frac{\text{d} f_{2}}{\text{d ln}p} -(\sigma + \frac{\sigma -3}{exp{[\frac{u_1a}{D(p)}]} -1} -\frac{\alpha u_1a}{D(p)(exp[\frac{u_1a}{D(p)}]-1)})f_{2}=-\sigma
\end{equation}
  where \(\alpha = \partial \ln D(p)/\partial \ln p\).

Although the two solutions above have complicated forms, substituting Eq.~\ref{equ:aj1} into Eq.~\ref{equ:aj2} allows one to obtain the expression for \(f_1(p)f_2(p)\):
\begin{equation}
f_1(p) f_2(p)
= C - \sigma \int_{p}^{\infty} f_1(p') \mathrm{d}\ln p'     
\end{equation}

Since\(f_1(p)f_2(p)\) must vanish as \(p\rightarrow \infty\), the downstream solution \(f_0(p)\) can be expressed as
\begin{equation}
f_0(p)
= - f_0 \sigma \int_{p}^{\infty} f_1(p') \mathrm{d}\ln p'    
\end{equation}
where \(f_0\) is a constant.

\vspace{5cm}
\nocite{*}

\bibliography{apssamp}

@article{manchester2005australia,
  title={The Australia telescope national facility pulsar catalogue},
  author={Manchester, R Nꎬ and Hobbs, G Bꎬ and Teoh, Aꎬ and Hobbs, M\_},
  journal={The Astronomical Journal},
  volume={129},
  number={4},
  pages={1993},
  year={2005},
  publisher={IOP Publishing}
}

@article{Hoshino:1992zz,
    author = "Hoshino, Masahiro and Arons, Jonathan and Gallant, Yves A. and Langdon, A. B.",
    title = "{Relativistic magnetosonic shock waves in synchrotron sources - Shock structure and nonthermal acceleration of positrons}",
    doi = "10.1086/171296",
    journal = "Astrophys. J.",
    volume = "390",
    pages = "454--479",
    year = "1992"
}

@article{reynolds2008supernova,
  title={Supernova remnants at high energy},
  author={Reynolds, Stephen P},
  journal={Annu. Rev. Astron. Astrophys.},
  volume={46},
  number={1},
  pages={89--126},
  year={2008},
  publisher={Annual Reviews}
}

@article{Xi:2018mii,
    author = "Xi, Shao-Qiang and Liu, Ruo-Yu and Huang, Zhi-Qiu and Fang, Kun and Wang, Xiang-Yu",
    title = "{GeV observations of the extended pulsar wind nebulae constrain the pulsar interpretations of the cosmic-ray positron excess}",
    eprint = "1810.10928",
    archivePrefix = "arXiv",
    primaryClass = "astro-ph.HE",
    doi = "10.3847/1538-4357/ab20c9",
    journal = "Astrophys. J.",
    volume = "878",
    number = "2",
    pages = "104",
    year = "2019"
}

@article{hillas2005can,
  title={Can diffusive shock acceleration in supernova remnants account for high-energy galactic cosmic rays?},
  author={Hillas, AM},
  journal={Journal of Physics G: Nuclear and Particle Physics},
  volume={31},
  number={5},
  pages={R95},
  year={2005},
  publisher={IOP Publishing}
}

@incollection{amato2024particle,
  title={Particle acceleration in pulsars and pulsar wind nebulae},
  author={Amato, Elena},
  booktitle={Foundations of Cosmic Ray Astrophysics},
  pages={183--221},
  year={2024},
  publisher={IOS Press}
}

@article{kroon2016electron,
  title={Electron acceleration in pulsar-wind termination shocks: an application to the Crab nebula gamma-ray flares},
  author={Kroon, John J and Becker, Peter A and Finke, Justin D and Dermer, Charles D},
  journal={The Astrophysical Journal},
  volume={833},
  number={2},
  pages={157},
  year={2016},
  publisher={IOP Publishing}
}

@article{hui2017rapid,
  title={Rapid X-ray variations of the Geminga pulsar wind nebula},
  author={Hui, CY and Lee, Jongsu and Kong, AKH and Tam, PHT and Takata, J and Cheng, KS and Ryu, Dongsu},
  journal={The Astrophysical Journal},
  volume={846},
  number={2},
  pages={116},
  year={2017},
  publisher={IOP Publishing}
}

@article{pavlov2010new,
  title={New X-ray observations of the Geminga pulsar wind nebula},
  author={Pavlov, George G and Bhattacharyya, Sudip and Zavlin, Vyacheslav E},
  journal={The Astrophysical Journal},
  volume={715},
  number={1},
  pages={66},
  year={2010},
  publisher={IOP Publishing}
}

@inproceedings{faherty2007trigonometric,
  title={The trigonometric parallax of the neutron star Geminga},
  author={Faherty, Jacqueline and Walter, Frederick M and Anderson, Jay},
  booktitle={Isolated Neutron Stars: From the Surface to the Interior},
  pages={225--230},
  year={2007},
  organization={Springer}
}

@article{abdo2007tev,
  title={TeV gamma-ray sources from a survey of the galactic plane with Milagro},
  author={Abdo, Aous A and Allen, B and Berley, D and Casanova, S and Chen, C and Coyne, DG and Dingus, BL and Ellsworth, RW and Fleysher, L and Fleysher, R and others},
  journal={The Astrophysical Journal},
  volume={664},
  number={2},
  pages={L91},
  year={2007},
  publisher={IOP Publishing}
}

@article{cao2019large,
  title={The large high altitude air shower observatory (LHAASO) science book (2021 edition)},
  author={Cao, Zhen and della Volpe, D and Liu, Siming and Bi, Xiaojun and Chen, Yang and Piazzoli, BD'Ettorre and Feng, Li and Jia, Huanyu and Li, Zhuo and Ma, Xinhua and others},
  journal={arXiv preprint arXiv:1905.02773},
  year={2019}
}

@article{manconi2024geminga,
  title={Geminga’s pulsar halo: An X-ray view},
  author={Manconi, Silvia and Woo, Jooyun and Shang, Ruo-Yu and Krivonos, Roman and Tang, Claudia and Di Mauro, Mattia and Donato, Fiorenza and Mori, Kaya and Hailey, Charles J},
  journal={Astronomy \& Astrophysics},
  volume={689},
  pages={A326},
  year={2024},
  publisher={EDP Sciences}
}

@article{amato2011streaming,
  title={The streaming instability: a review},
  author={Amato, Elena},
  journal={Memorie della Societa Astronomica Italiana, v. 82, p. 806 (2011)},
  volume={82},
  pages={806},
  year={2011}
}

@article{strong2007cosmic,
  title={Cosmic-ray propagation and interactions in the Galaxy},
  author={Strong, Andrew W and Moskalenko, Igor V and Ptuskin, Vladimir S},
  journal={Annu. Rev. Nucl. Part. Sci.},
  volume={57},
  number={1},
  pages={285--327},
  year={2007},
  publisher={Annual Reviews}
}

@article{guo2021observations,
  title={Observations of extended very-high-energy halos around Geminga and Monogem with the LHAASO-KM2A},
  author={Guo, Yingying and Zhang, Yi and Yuan, Qiang and Hu, Hongbo and Cao, Zhen and Aharonian, F and An, Q and Bai, LX and Bai, YX and Bao, YW and others},
  journal={PoS ICRC2021},
  volume={964},
  year={2021}
}

@article{knies2018suzaku,
  title={Suzaku observations of the Monogem Ring and the origin of the Gemini H $\alpha$ ring},
  author={Knies, Jonathan R and Sasaki, Manami and Plucinsky, Paul P},
  journal={Monthly Notices of the Royal Astronomical Society},
  volume={477},
  number={4},
  pages={4414--4422},
  year={2018},
  publisher={Oxford University Press}
}

@article{pellizzoni2011detection,
  title={Detection of continuum radio emission associated with Geminga},
  author={Pellizzoni, A and Govoni, F and Esposito, P and Murgia, M and Possenti, A},
  journal={Monthly Notices of the Royal Astronomical Society: Letters},
  volume={416},
  number={1},
  pages={L45--L49},
  year={2011},
  publisher={The Royal Astronomical Society}
}

@article{Bykov:2017xpo,
    author = "Bykov, A. M. and Amato, E. and Petrov, A. E. and Krassilchtchikov, A. M. and Levenfish, K. P.",
    title = "{Pulsar wind nebulae with bow shocks: non-thermal radiation and cosmic ray leptons}",
    eprint = "1705.00950",
    archivePrefix = "arXiv",
    primaryClass = "astro-ph.HE",
    doi = "10.1007/s11214-017-0371-7",
    journal = "Space Sci. Rev.",
    volume = "207",
    number = "1-4",
    pages = "235--290",
    year = "2017"
}

@article{Posselt_2017,
   title={GEMINGA’S PUZZLING PULSAR WIND NEBULA},
   volume={835},
   ISSN={1538-4357},
   url={http://dx.doi.org/10.3847/1538-4357/835/1/66},
   DOI={10.3847/1538-4357/835/1/66},
   number={1},
   journal={The Astrophysical Journal},
   publisher={American Astronomical Society},
   author={Posselt, B. and Pavlov, G. G. and Slane, P. O. and Romani, R. and Bucciantini, N. and Bykov, A. M. and Kargaltsev, O. and Weisskopf, M. C. and Ng, C.-Y.},
   year={2017},
   month=jan, pages={66} }

@article{dempsey2007radiative,
  title={Radiative losses and cut-offs of energetic particles at relativistic shocks},
  author={Dempsey, Paul and Duffy, Peter},
  journal={Monthly Notices of the Royal Astronomical Society},
  volume={378},
  number={2},
  pages={625--634},
  year={2007},
  publisher={Blackwell Publishing Ltd Oxford, UK}
}

@article{kirk2000particle,
  title={Particle acceleration at ultrarelativistic shocks: an eigenfunction method},
  author={Kirk, JG and Guthmann, AW and Gallant, YA and Achterberg, A},
  journal={The Astrophysical Journal},
  volume={542},
  number={1},
  pages={235},
  year={2000},
  publisher={IOP Publishing}
}

@article{Foreman-Mackey_2013,
doi = {10.1086/670067},
url = {https://dx.doi.org/10.1086/670067},
year = {2013},
month = {feb},
publisher = {University of Chicago Press},
volume = {125},
number = {925},
pages = {306},
author = {Daniel Foreman-Mackey and David W. Hogg and Dustin Lang and Jonathan Goodman},
title = {emcee: The MCMC Hammer},
journal = {Publications of the Astronomical Society of the Pacific}
}

@article{yuan2017propagation,
  title={Propagation of cosmic rays in the AMS-02 era},
  author={Yuan, Qiang and Lin, Su-Jie and Fang, Kun and Bi, Xiao-Jun},
  journal={Physical Review D},
  volume={95},
  number={8},
  pages={083007},
  year={2017},
  publisher={APS}
}

@article{zabalza2015naima,
  title={naima: a Python package for inference of relativistic particle energy distributions from observed nonthermal spectra},
  author={Zabalza, V{\'\i}ctor},
  journal={arXiv preprint arXiv:1509.03319},
  year={2015}
}

@article{fang2021klein,
  title={Klein--Nishina effect and the cosmic ray electron spectrum},
  author={Fang, Kun and Bi, Xiao-Jun and Lin, Su-Jie and Yuan, Qiang},
  journal={Chinese Physics Letters},
  volume={38},
  number={3},
  pages={039801},
  year={2021},
  publisher={IOP Publishing}
}

@article{florinski2013galactic,
  title={Galactic cosmic rays in the outer heliosphere: Theory and models},
  author={Florinski, V and Ferreira, SES and Pogorelov, NV},
  journal={Space Science Reviews},
  volume={176},
  pages={147--163},
  year={2013},
  publisher={Springer}
}

@article{florinski2003galactic,
  title={Galactic cosmic ray transport in the global heliosphere},
  author={Florinski, V and Zank, GP and Pogorelov, NV},
  journal={Journal of Geophysical Research: Space Physics},
  volume={108},
  number={A6},
  year={2003},
  publisher={Wiley Online Library}
}

@article{kota2013theory,
  title={Theory and modeling of galactic cosmic rays: Trends and prospects},
  author={K{\'o}ta, J{\'o}zsef},
  journal={Space Science Reviews},
  volume={176},
  pages={391--403},
  year={2013},
  publisher={Springer}
}

@book{stanev2010high,
  title={High energy cosmic rays},
  author={Stanev, Todor and Stanev},
  volume={175},
  year={2010},
  publisher={Springer}
}

@article{florinski2003cosmic,
  title={Cosmic-ray spectra at spherical termination shocks},
  author={Florinski, V and Jokipii, JR},
  journal={The Astrophysical Journal},
  volume={591},
  number={1},
  pages={454},
  year={2003},
  publisher={IOP Publishing}
}

@article{heavens1987particle,
  title={Particle acceleration in extragalactic sources: the role of synchrotron losses in determining the spectrum},
  author={Heavens, AF and Meisenheimer, K},
  journal={Monthly Notices of the Royal Astronomical Society},
  volume={225},
  number={2},
  pages={335--353},
  year={1987},
  publisher={Oxford University Press}
}

@article{sironi2017particle,
  title={Particle acceleration in pulsar wind nebulae: PIC modelling},
  author={Sironi, Lorenzo and Cerutti, Beno{\^\i}t},
  journal={Modelling Pulsar Wind Nebulae},
  pages={247--277},
  year={2017},
  publisher={Springer}
}

@article{petri2008magnetic,
  title={Magnetic reconnection at the termination shock of a striped pulsar wind},
  author={P{\'e}tri, J{\'e}r{\^o}me and Lyubarsky, Yuri},
  journal={International Journal of Modern Physics D},
  volume={17},
  number={10},
  pages={1961--1967},
  year={2008},
  publisher={World Scientific}
}

@article{sironi2011acceleration,
  title={Acceleration of particles at the termination shock of a relativistic striped wind},
  author={Sironi, Lorenzo and Spitkovsky, Anatoly},
  journal={The Astrophysical Journal},
  volume={741},
  number={1},
  pages={39},
  year={2011},
  publisher={IOP Publishing}
}

@article{drury1983introduction,
  title={An introduction to the theory of diffusive shock acceleration of energetic particles in tenuous plasmas},
  author={Drury, L O'C},
  journal={Reports on Progress in Physics},
  volume={46},
  number={8},
  pages={973},
  year={1983},
  publisher={IOP Publishing}
}

@article{caprioli2009escape,
  title={On the escape of particles from cosmic ray modified shocks},
  author={Caprioli, D and Blasi, P and Amato, E},
  journal={Monthly Notices of the Royal Astronomical Society},
  volume={396},
  number={4},
  pages={2065--2073},
  year={2009},
  publisher={The Royal Astronomical Society}
}

@article{di2019detection,
  title={Detection of a $\gamma$-ray halo around Geminga with the Fermi-LAT data and implications for the positron flux},
  author={Di Mauro, Mattia and Manconi, Silvia and Donato, Fiorenza},
  journal={Physical Review D},
  volume={100},
  number={12},
  pages={123015},
  year={2019},
  publisher={APS}
}

@article{klecker1998anomalous,
  title={Anomalous cosmic rays},
  author={Klecker, B and Mewaldt, RA and Bieber, JW and Cummings, AC and Drury, L and Giacalone, J and Jokipii, JR and Jones, FC and Krainev, MB and Lee, MA and others},
  journal={Space Science Reviews},
  volume={83},
  pages={259--308},
  year={1998},
  publisher={Springer}
}

@article{xi2019gev,
  title={GeV observations of the extended pulsar wind nebulae constrain the pulsar interpretations of the cosmic-ray positron excess},
  author={Xi, Shao-Qiang and Liu, Ruo-Yu and Huang, Zhi-Qiu and Fang, Kun and Wang, Xiang-Yu},
  journal={The Astrophysical Journal},
  volume={878},
  number={2},
  pages={104},
  year={2019},
  publisher={IOP Publishing}
}

@article{albert2024precise,
  title={Precise Measurements of TeV Halos around Geminga and Monogem Pulsars with HAWC},
  author={Albert, A and Alfaro, R and Alvarez, C and Arteaga-Vel{\'a}zquez, JC and Rojas, D Avila and Solares, HA Ayala and Babu, R and Belmont-Moreno, E and Bernal, A and Caballero-Mora, KS and others},
  journal={The Astrophysical Journal},
  volume={974},
  number={2},
  pages={246},
  year={2024},
  publisher={IOP Publishing}
}

@article{abeysekara2017extended,
  title={Extended gamma-ray sources around pulsars constrain the origin of the positron flux at Earth},
  author={Abeysekara, AU and Albert, Andrea and Alfaro, R and Alvarez, C and {\'A}lvarez, JD and Arceo, R and Arteaga-Vel{\'a}zquez, JC and Avila Rojas, D and Ayala Solares, HA and Barber, AS and others},
  journal={Science},
  volume={358},
  number={6365},
  pages={911--914},
  year={2017},
  publisher={American Association for the Advancement of Science}
}

@article{caraveo2003geminga,
  title={Geminga's tails: A pulsar bow shock probing the interstellar medium},
  author={Caraveo, PA and Bignami, GF and DeLuca, A and Mereghetti, S and Pellizzoni, A and Mignani, R and Tur, A and Becker, W},
  journal={Science},
  volume={301},
  number={5638},
  pages={1345--1347},
  year={2003},
  publisher={American Association for the Advancement of Science}
}

@article{caraveo1996parallax,
  title={Parallax observations with the Hubble Space Telescope yield the distance to Geminga},
  author={Caraveo, Patrizia A and Bignami, Giovanni F and Mignani, Roberto and Taff, Laurence G},
  journal={The Astrophysical Journal},
  volume={461},
  number={2},
  pages={L91},
  year={1996},
  publisher={IOP Publishing}
}

@article{Mukhopadhyay:2021dyh,
    author = "Mukhopadhyay, Payel and Linden, Tim",
    title = "{Self-generated cosmic-ray turbulence can explain the morphology of TeV halos}",
    eprint = "2111.01143",
    archivePrefix = "arXiv",
    primaryClass = "astro-ph.HE",
    doi = "10.1103/PhysRevD.105.123008",
    journal = "Phys. Rev. D",
    volume = "105",
    number = "12",
    pages = "123008",
    year = "2022"
}

@article{evoli2018self,
  title={Self-generated cosmic-ray confinement in TeV halos: Implications for TeV $\gamma$-ray emission and the positron excess},
  author={Evoli, Carmelo and Linden, Tim and Morlino, Giovanni},
  journal={Physical Review D},
  volume={98},
  number={6},
  pages={063017},
  year={2018},
  publisher={APS}
}

@article{fang2023effect,
  title={Effect of the magnetic field correlation length on the gamma-ray pulsar halo morphology under anisotropic diffusion},
  author={Fang, Kun and Hu, Hong-Bo and Bi, Xiao-Jun and Chen, En-Sheng},
  journal={Physical Review D},
  volume={108},
  number={2},
  pages={023017},
  year={2023},
  publisher={APS}
}

@article{liu2019understanding,
  title={Understanding the multiwavelength observation of Geminga’s TeV halo: the role of anisotropic diffusion of particles},
  author={Liu, Ruo-Yu and Yan, Huirong and Zhang, Heshou},
  journal={Physical review letters},
  volume={123},
  number={22},
  pages={221103},
  year={2019},
  publisher={APS}
}

@article{giacinti2020halo,
  title={Halo fraction in TeV-bright pulsar wind nebulae},
  author={Giacinti, G and Mitchell, AMW and L{\'o}pez-Coto, R and Joshi, V and Parsons, RD and Hinton, JA},
  journal={Astronomy \& Astrophysics},
  volume={636},
  pages={A113},
  year={2020},
  publisher={EDP Sciences}
}

@article{blumenthal1970bremsstrahlung,
  title={Bremsstrahlung, synchrotron radiation, and compton scattering of high-energy electrons traversing dilute gases},
  author={Blumenthal, George R and Gould, Robert J},
  journal={Reviews of modern Physics},
  volume={42},
  number={2},
  pages={237},
  year={1970},
  publisher={APS}
}

@article{fang2019possible,
  title={Possible origin of the slow-diffusion region around Geminga},
  author={Fang, Kun and Bi, Xiao-Jun and Yin, Peng-Fei},
  journal={Monthly Notices of the Royal Astronomical Society},
  volume={488},
  number={3},
  pages={4074--4080},
  year={2019},
  publisher={Oxford University Press}
}

\end{document}